\documentclass[aps,tightenlines,nofootinbib,preprint,groupedaddress]{revtex4}
\usepackage{graphicx}
\newcommand{\be}{\begin{equation}}
\newcommand{\ee}{\end{equation}}
\newcommand{\bear}{\begin{eqnarray}}
\newcommand{\eear}{\end{eqnarray}}
\begin{document}

\title{Surviving the renormalon in heavy quark
potential}

\author{Taekoon Lee}
\email{tlee@muon.kaist.ac.kr}

\affiliation{Department of Physics, Korea Advanced Institute of
Science and Technology, Daejon 305-701, Korea}

%\date{\today}

\begin{abstract}
We show  that the Borel resummed 
perturbative static potential at $N_f=0$ converges well, and is
in a remarkable agreement with the quenched lattice calculation at 
distances $1/r\agt 660$ MeV. This shows that Borel resummation is 
very good at handling the renormalon in the static potential (and in
the pole mass), and allows one to use the pole mass in perturbative
calculation of heavy quark physics. 

\end{abstract}


%\keywords{}

\maketitle

\section{\label{sec1}Introduction}
\begin{figure}
 \includegraphics[angle=-90 , width=10cm
 ]{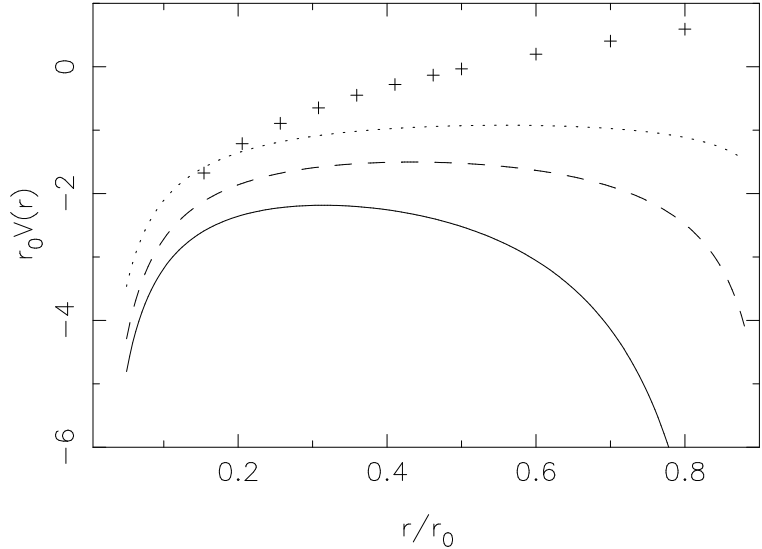}
\caption{\label{fig1} % \footnotesize
The static potential at leading order (dotted), next-leading order (dashed),
and next-next-leading order (solid).
The data points denote lattice potential.}
\end{figure}

The asymptotic freedom of quantum chromodynamics (QCD) allows one to calculate
the short distance physics accurately using perturbation. Unexpectedly,
however, the perturbative expansion of the static potential between
a quark-antiquark pair does not show a convergence even at very short
distances (see Fig. \ref{fig1}). Moreover, no agreement is seen with
the accurate lattice calculations of the static potential.
This led to a suggestion of nonperturbative linear potential at
short distance \cite{bali1},
which, if proven true, would violate the expectation
of the operator product expansion (OPE) that the nonperturbative effect
at short distance is at most a quadratic potential.

On the other hand, the bad convergence 
behavior of the perturbative expansion of the
potential is well understood to be caused by the infrared (IR) renormalon
which induces a constant nonperturbative
effect proportional to  $\Lambda_{\rm QCD}$ \cite{al}.
This prompted several approaches to the problem.
One is based on the observation that the force between a pair of 
static quarks is free from the leading 
renormalon. The potential obtained 
by integrating the force calculated in perturbation indeed agrees 
quite well 
at short distance with the lattice potential \cite{sommer1}, up to an
$r$ independent constant.
Another approach is the renormalon subtracted (RS) scheme \cite{pineda1},
in  which one subtracts order by order the renormalon contribution
from the perturbative potential. The potential calculated
in this way also shows an improved convergence and agreement
with the lattice potential.
Another idea is to employ the cancellation of the renormalons in the
static potential and the pole mass of the heavy quark \cite{hoang,beneke-98}.
By expanding the pole mass and the static potential of a color singlet
quarkonium in the running coupling $\alpha_s(\mu)$ 
and a short distance mass $m(\mu)$ one can avoid the renormalon problem,
and indeed such an  
expansion shows an improved convergence 
\cite{sumino1,sumino-pt1,sumino-pt2}.

In this paper we show a more direct
approach to the problem is possible
via the  Borel resummation of the perturbative potential.
Since one might believe that the presence of an IR renormalon 
makes Borel resummation impossible, we state in advance that
it is perfectly possible in this case.
An IR renormalon in Borel resummation merely demands a corresponding
nonperturbative effect, and since in this case it is a constant,
the $r$ dependence of the potential can be resummed with no difficulty.
Moreover, this renormalon caused  nonperturbative effect
could be computed in the framework introduced in \cite{lee1}
where the nonperturbative effect is determined based on its conjectured
analyticity in the complex coupling plane.
An obvious advantage of the direct resummation is
that the  normalization of the potential can be fixed. 
In the approaches based on the renormalon cancellation/absence
the potential can be fixed only up to an $r$ independent constant.

As we shall see the Borel resummed potential at short distance
converges quickly, and agrees remarkably well 
with the lattice calculation, in fact better than any other approach
introduced so far. The implication of this
is significant. In the perturbative 
calculation of a heavy quark system one does not have to give up
the pole mass in favor of a short distance mass
to avoid the renormalon problem, 
and still can have a tight control on
the perturbative expansion.

Throughout the paper, unless stated otherwise, we consider pure
QCD with no active quark flavors ($N_f=0$), and the perturbative
expansions considered are assumed to be in the
$\overline{\rm MS}$ scheme. As for the renormalon, we restrict 
our attention to the leading infrared renormalon that is closest
to the origin in the Borel plane.

\section{\label{sec2}Bilocal expansion of the Borel transform}

In general the perturbative expansion in weak coupling constant is an
asymptotic expansion. When the large order behavior of the expansion
is sign alternating like in $\phi^4$ theory it may be Borel resummed.
However, when the expansion is of same sign at large orders Borel
resummation demands a more careful treatment \cite{justin0}.
In the case of the latter,
one can first do Borel resummation at an unphysical negative coupling,
at which the series is sign alternating,
and then do analytic continuation in the complex coupling plane
to the physical positive coupling.
The Borel resummed amplitude obtained in such a way, however, 
turns out to have a cut along
the positive real axis in the coupling plane, and consequently has
an ambiguous imaginary part at a physical coupling.
In Borel integration this imaginary part arises
precisely from the infrared (IR) renormalon singularity 
of the Borel transform on the integration contour.
This unphysical, ambiguous imaginary part then must be canceled by
the nonperturbative effect corresponding to the renormalon. For further
details we refer to \cite{lee1}.

Thus the static inter-quark potential V(r)\footnote{Because of its infrared
sensitivity the static potential is dependent on the ultrasoft factorization
scale beginning at NNNLO \cite{ps1,ps2}, however, to the order we are
concerned (NNLO) this can be ignored.},
which has an IR renormalon,
can be written as  the sum of the Borel integration with a 
contour on the upper (or lower) half plane and the nonperturbative effect
\cite{lee1},
\be
V[r,\alpha_s(1/r)\pm i\epsilon]=\frac{1}{r\beta_0}
\int_{0\pm i\epsilon}^{
\infty\pm i\epsilon}
e^{-b/\beta_0\alpha_s(1/r)}\tilde V (b)\, d b
+ V_{\rm NP}[r,\alpha_s(1/r)\pm i\epsilon]\,
\label{e1}
\ee
where $\beta_0$ is the one loop coefficient of the QCD $\beta$ function,
\bear
\beta(\alpha_s)&=&\mu^2\frac{d\alpha_s}{d \mu^2} \nonumber \\
&=&-\alpha_s^2(\beta_0 +\beta_1 \alpha_s +\beta_2 \alpha_s^2+{\ldots} )\,,
\label{e2}
\eear
and
$\tilde V(b)$ is the Borel transform that
is given by
\be
\tilde V(b)=\sum_{n=0}^{\infty} \frac{V_n}{n!}
\left(\frac{b}{\beta_0}\right)^n \,,
\label{e3}
\ee
with $V_n$ defined in the perturbative expansion of the potential,
\be
V(r,\alpha_s)=\frac{1}{r}\sum_n^\infty V_n \alpha_s^{n+1} \,.
\label{e4}
\ee
$V_{\rm NP}$ denotes the renormalon caused nonperturbative effect.
Since the imaginary parts in the first term in Eq. (\ref{e1})
and in $V_{\rm NP}$,
respectively, cancel, the potential can be written as
\be
V[r,\alpha_s(1/r)]=\frac{1}{r\beta_0}{\rm Re}\left[\int_{0\pm i\epsilon}^{
\infty\pm i\epsilon}
e^{-b/\beta_0\alpha_s(1/r)}\tilde V (b)\, d b\right]
+ {\rm Re}\left\{ V_{\rm NP}[r,\alpha_s(1/r)\pm i\epsilon]\right\}\,.
\label{e5}
\ee
Since $V_{\rm NP}$ is an $r$ independent constant proportional to
$\Lambda_{\rm QCD}$
we can ignore it as far as the $r$ dependence of the potential
is concerned. However, a discussion on its determination will be 
given later on.

The cancellation of the imaginary parts in the integral term and $V_{\rm NP}$
in Eq. (\ref{e1})
determines the renormalon singularity in the Borel transform
$\tilde V(b)$.
By comparing the functional form of 
\bear
V_{\rm NP} &\propto& \Lambda_{\overline{\rm MS}} \nonumber\\
&\propto& \frac{1}{r}\alpha_s(1/r)^{-\nu}
e^{-1/2\beta_0\alpha_s(1/r)} \left[1
-\frac{1}{2}(\beta_2\beta_0-\beta_1^2)/\beta_0^3\alpha_s(1/r)
+{\ldots} \right]
\eear
 with the
imaginary part of the Borel integration
term in (\ref{e1}), one can see $\tilde V(b)$ must have the singularity
\bear
\tilde V(b) = \frac{c_V}{(1-2b)^{1+\nu}}\left[
1+c_1 (1-2b)+c_2(1-2b)^2+{\ldots}\right] + \text{ Analytic part}\,,
\label{e6}
\eear
where
the ``Analytic part'' denotes terms analytic around $b=1/2$.
The constants $\nu$ and $c_i$, which depend only on the
coefficients of the  $\beta$ function, were first determined 
in \cite{beneke1}, and can be computed up to
$c_2$ from the known four loop $\beta$ function \cite{betafunction}:
\bear
\nu &=& \frac{\beta_1}{2\beta_0^2}\,, \hspace{.25in}
c_1=\frac{
\beta_1^2-\beta_0\beta_2}{4\nu\beta_0^4}\,, \nonumber\\
c_2&=& \frac{
 \beta_1^4 +4\beta_0^3\beta_1\beta_2
-2 \beta_0\beta_1^2\beta_2
+\beta_0^2(\beta_2^2-2\beta_1^3)-2\beta_3\beta_0^4}{
32\nu(\nu-1)\beta_0^8} \,.
\eear

The residue $c_V$ becomes the normalization constant  of 
the large order behavior of the expansion (\ref{e4}), and its exact value is
not known, but it can be determined perturbatively using the method
developed in \cite{lee2,lee3}.
Once $c_V$ is known, we can combine the two expansions of the Borel transform
(\ref{e3}) and (\ref{e6}) at $b=0$ and at $b=1/2$, respectively, to obtain an
improved description of the Borel transform in the region
between the origin and the renormalon location at $b=1/2$.
There are in principle an infinite number of ways to interpolate the two
expansions, but here we shall take a simple one which turns out to suffice
our purpose very well. We write the Borel transform as a two
point expansion, which we  call a {\it bilocal expansion} \footnote{
This was first introduced in \cite{lee4} in a slightly different context.}:
\bear
\tilde V(b) &=&\lim_{N,M \to \infty} \tilde V_{\rm N,M} (b) \nonumber\\
&=& \lim_{N,M \to \infty}\left\{
\sum_{n=0}^N\frac{h_n}{n!} \left(\frac{b}{\beta_0}\right)^n
+\frac{c_V}{(1-2b)^{1+\nu}}\left[ 1 +\sum_{i=1}^M c_i (1-2b)^i\right]\right\}
\label{e8} \,.
\eear
By demanding that this bilocal expansion reproduce the expansion (\ref{e3})
around the origin  
the coefficients $h_n$ can be determined in terms of
$V_n$ and $c_i$. This gives, for example, the first three coefficients as
\bear
h_0&=& V_0 -c_V(1+c_1+c_2) \,, \nonumber\\
h_1&=&V_1 -2c_V\beta_0[1-c_2+\nu(1+c_1+c_2)] \,,\nonumber\\
h_2&=& V_2-4c_V\beta_0^2[2+\nu(3+c_1-c_2)+\nu^2(1+c_1+c_2)]\,.
\eear
For the bilocal expansion to work it is essential to have 
the residue $c_V$ calculated in a good accuracy, which is the subject
of the next section.

\section{Renormalon residue}
The residue can be determined in perturbation using the method 
developed in \cite{lee2,lee3}. It was shown in \cite{pineda2,pineda3,pineda1} 
that the residue in the case of the static potential
can be calculated quite accurately.
For completeness, we repeat the calculation here, and 
in the meantime obtain an improved estimate.

To compute $c_V$ we first consider the function
\be
R(b)\equiv (1-2b)^{1+\nu} \tilde V(b)\,.
\ee
Then,
\be
c_V=R(\frac{1}{2})\,.
\ee
$R(b)$ has a cut, but is bounded, at $b=1/2$, and thus we can 
write $c_V$ as a convergent series,
\be
c_V=\sum_{n=0}^\infty r_n \left(\frac{1}{2}\right)^n\,,
\ee
where $r_n$ are the coefficients of the power expansion of
$R(b)$ at the origin.
The first three $r_n$ can be calculated from the known 
$V_n$ up to next-next-leading order (NNLO) \cite{fischler,peter,schroder},
and this gives
\be
c_V\approx-1.33333 +0.49943-0.33844=-1.17234\,.
\ee
The convergence is not that rapid but the series is oscillating.
An important observation made in \cite{pineda2}
is that the reliability of this estimate can be checked by the
mutual cancellation of the renormalons in the static potential
and the  pole mass.

In perturbation theory the
heavy quark pole mass $m_{\rm pole}$ can be expanded as
\be
m_{\rm pole}[\alpha_s(m_{\overline {\rm MS}})]=
m_{\overline {\rm MS}}\left[
1+ \sum_{n=0}^{\infty}
p_n \alpha_s(m_{\overline {\rm MS}})^{n+1}\right] \,,
\label{e14}
\ee
where  $m_{\overline{\rm MS}}$ [$\equiv
m_{\overline {\rm MS}}(m_{\overline {\rm MS}})$] denotes the 
$\overline {\rm MS}$ mass.
As in the case of the static potential the Borel resummed pole mass 
can be written as
\bear
m_{\rm pole}[\alpha_s(m_{\overline {\rm MS}})\pm i\epsilon]
=&& m_{\overline {\rm MS}}\left[ 1 +\frac{1}{\beta_0}
\int_{0\pm i\epsilon}^{\infty \pm i\epsilon}
e^{-b/\beta_0\alpha_s(m_{\overline {\rm MS}}) } \tilde m_{\rm pole}
(b) \,db\right]
\nonumber \\
&&+ m_{\rm NP}[\alpha_s(m_{\overline {\rm MS}}) \pm i\epsilon]\,,
\label{e15}
\eear
where the Borel transform $\tilde m_{\rm pole}(b)$ has
the perturbative expansion
\be
\tilde m_{\rm pole}(b)=\sum_{n=0}^\infty \frac{p_n}{n!}
\left(\frac{b}{\beta_0}\right)^n\,,
\ee
and $m_{\rm NP}$ denotes the renormalon induced nonperturbative effect.
The renormalon ambiguity in the pole mass proportional to
$\Lambda_{\overline{\rm MS}}$ gives rise to a renormalon singularity that
has exactly the same form as Eq. (\ref{e6}) of the static potential, 
\bear
\tilde m_{\rm pole}(b) = \frac{c_m}{(1-2b)^{1+\nu}}\left[
1+c_1 (1-2b)+c_2(1-2b)^2+{\ldots} \right] + \text{ Analytic part}\,.
\eear
Now the 
cancellation of the renormalons in $2m_{\rm pole}$ and
$ V(r)$ \cite{hoang,beneke-98} leads to
\be
c_V+2c_m=0\,.
\label{e18}
\ee

We shall now compute the residue $c_m$ following the computation of
$c_V$. Using the known coefficients up to NNLO
\cite{massexpansion1,massexpansion2,massexpansion3}
of the expansion (\ref{e14}) we have
\be
c_m \approx 0.42441+ 0.17473+0.02289=0.62203
\label{e19}
\ee
This time the convergence is quite good.
With the two computed values we now have
\be
\frac{c_V+2c_m}{c_V-2c_m}= 0.02968 \,,
\label{e20}
\ee
which shows a remarkable cancellation of the two residues. This
gives an assurance on the accuracy of the 
calculated residues.

We shall now  compute $c_m$ in a slightly different way.
As has been shown in solvable models \cite{lee1}, the knowledge on the
 renormalon locations in the Borel plane
can be used in improving the  residue calculation.
Since we are interested in the power expansion of $R(b)$ around the origin,
we can obtain in principle a better convergence 
by expanding it in a new complex plane in which it is 
smoother around the origin \cite{lee4}. This can be done by pushing 
the renormalon singularities save the first one away from the 
origin with a 
conformal mapping. Let us consider the mapping \cite{lee4,lee5}
\be
w= \frac{\sqrt{1+b}-\sqrt{1-2b/3}}{\sqrt{1+b}+\sqrt{1-2b/3}}\,,
\label{e21}
\ee
which maps the first renormalon at $b=1/2$ to $w=w_0$, where 
\be
w_0= \frac{1}{5}\,,
\ee
and all other renormalons (at $b=-n$ and $b=1/2+n$ where $n=1,2,3, {\ldots}$) 
onto the unit circle.

Expanding  $R[b(w)]$ at the origin to $O(w^2)$ and evaluating it
at $w=w_0$ we have a new estimate of $c_m$
\be
c_m \approx 0.42441 +0.16774+0.03451=0.62667\,,
\label{e23}
\ee
which is quite close to the previous one (\ref{e19}). This stability is
reassuring that our computation is accurate.

Now we shall quantitatively
estimate the error in the computed residue (\ref{e23}).
We do this by computing $c_m$ using an estimated NNNLO coefficient of the
expansion (\ref{e14}).
We first estimate the unknown NNNLO coefficient $p_3$ following the 
method developed in \cite{lee5}.
First, expand $R[b(w)]$ to $O(w^3)$ with  $p_3$ included.
This gives
\be
R[b(w)]=0.42441+0.83872 w+0.86284 w^2+(-129.2687+3.43505 \,p_3) w^3 \,.
\ee
Note that the $p_3$-independent constant term in the coefficient
of $w^3$ is much larger than the coefficients of the lower orders.
It turns out this is a generic feature of an asymptotic expansion
with rapidly growing coefficients, and it can be used in estimating 
higher order unknown coefficients. From the pattern of the known 
lower order terms
it appears quite reasonable to assume that 
the fourth coefficient is bounded by
\be
|129.2687-3.43505\,p_3|<2 \,.
\ee
This gives an estimate on $p_3$
\be
          p_3= 37.6322\pm 0.58223 \,.
\ee
With this result we can repeat the computation
of $c_m$ in $w$ plane, now at NNNLO, to obtain
\be
c_m= 0.62667\pm 0.02553\,.
\label{e27}
\ee
We thus conclude the error in the computed residue (\ref{e23})
is about 4\%.

For the numerical analysis in Sec. \ref{sec5}
we use the exact relation (\ref{e18}) and the pole mass residue
(\ref{e23}) to compute $c_V$. Since the convergence 
in the calculation of the pole mass residue is better than
that of the potential, we would have a more accurate value 
this way. We thus have
\be
c_V= -1.25334 \pm 0.05106\,.
\label{e28}
\ee

\section{Determination of the nonperturbative effect}

In this section we give an evaluation of the renormalon caused
nonperturbative effect $V_{\rm NP}$ 
using the method developed in \cite{lee1}.
As mentioned in Sec. \ref{sec2}
the role of $V_{\rm NP}$ in Borel resummation
is to cancel the imaginary part arising from the renormalon singularity
in the Borel integration of the static potential.
This means that in principle the imaginary part of 
$V_{\rm NP}$ is calculable from perturbation theory.
However, the real part, which is the physical quantity, cannot
be directly calculated perturbatively without a further input.

The method for computing the real part 
relies on the  analyticity of $V_{\rm NP}$ in the
complex $\alpha_s$ plane.
As mentioned, 
the potential obtained by Borel resumming the asymptotic expansion
has a cut along the positive real
axis in the $\alpha_s$ plane, and from this cut the imaginary part of the
perturbative term, the integral term in Eq. (\ref{e1}),  arises.
To cancel this imaginary part 
it is thus plausible to demand that $V_{\rm NP}(r,\alpha_s)$ also
have a cut only along the positive real axis in the coupling plane.
This then relates the real part to the perturbatively 
calculable imaginary part (we refer to \cite{lee1} for details).
For convenience, we shall call this method of determining
the nonperturbative effect (along with the Borel integration of
the perturbation series) `analytic Borel resummation (ABR)'.
Some nonperturbative effects in solvable models were shown to be
calculable in ABR \cite{lee1}.

For ABR to work it is essential to have the functional form  of the
nonperturbative effect beforehand. In the case of the static
potential it is provided by the renormalization group equation.
Since $V_{\rm NP}$ in the $\overline{\rm MS}$ scheme 
should be a constant proportional to
$\Lambda_{\overline{\rm MS}}$, where  
\bear
\Lambda_{\overline{\rm MS}}= \frac{1}{r} [\beta_0 \alpha_s(1/r)]^{-\nu}
e^{-1/2\beta_0\alpha_s(1/r)} \exp \left\{-\frac{1}{2}
\int_0^{\alpha_s(1/r)}\left[
\frac{1}{\beta(x)}+\frac{1}{\beta_0 x^2}-
\frac{\beta_1}{\beta_0^2 x}\right]\,dx\right\}\,,
\label{lambdams}
\eear
we can write, by demanding $V_{\rm NP}$ have a cut only along the positive
real axis, 
\bear
V_{\rm NP}[r,\alpha_s(1/r)]=&&\frac{C}{r} [- \alpha_s(1/r)]^{-\nu}
e^{-1/2\beta_0\alpha_s(1/r)} \nonumber \\
&&\times \exp \left\{-\frac{1}{2}
\int_0^{\alpha_s(1/r)}\left[
\frac{1}{\beta(x)}+\frac{1}{\beta_0\alpha_s^2}-
\frac{\beta_1}{\beta_0^2 x}\right]\,dx\right\}\,,
\eear
with $C$ an undetermined real constant. Note that a cut can arise only from
the prefactor  in Eq. (\ref{lambdams}) with a noninteger $\nu$.
Now the cancellation of the imaginary part in $V_{\rm NP}[r,\alpha_s(1/r)
\pm i \epsilon]$ with the corresponding imaginary part in the
Borel integration term in Eq. (\ref{e1}) fixes the constant $C$:
\be
C=\frac{c_V \Gamma(-\nu)}{(2\beta_0)^{1+\nu}}\,.
\ee
The real part of $V_{\rm NP}$ is then given by
\bear
{\rm Re} \left[ V_{\rm NP}(\alpha_s \pm i\epsilon)\right]
= \frac{c_V \Gamma(-\nu)}{2^{1+\nu} \beta_0} \cos(\nu\pi) \Lambda_{\overline
{\rm MS}}\,.
\label{nonp_v}
\eear
With the calculated residue $c_V$ in  Eq. (\ref{e28}), we find at $N_f=0$
\be
{\rm Re} \left[ V_{\rm NP}(\alpha_s \pm i\epsilon)\right]
= 0.477 \Lambda_{\overline{\rm MS}}\,.
\label{e34}
\ee
In the numerical analysis in the next section we will combine
this result with the Borel integration of the perturbative
expansions.

\section{\label{sec5}Comparison with lattice calculation}

The static potential in lattice calculation is extracted from the 
Wilson line of a static quark-antiquark pair, computed in
Monte Carlo simulation. The recent calculations
\cite{lattice1,lattice2,lattice3,lattice4}
employing
large lattices up to $64^4$ achieved a remarkable accuracy, and 
can probe a short distance where perturbative QCD should be applicable.
It is thus an ideal place where  perturbative QCD can be
compared with lattice calculations.

As we mentioned in Introduction, the truncated power series of the
perturbative expansion fails even at very short distance.
We shall now see this problem can be cured by Borel resummation.

\begin{figure}
 \includegraphics[angle=-90 ,width=10cm
 ]{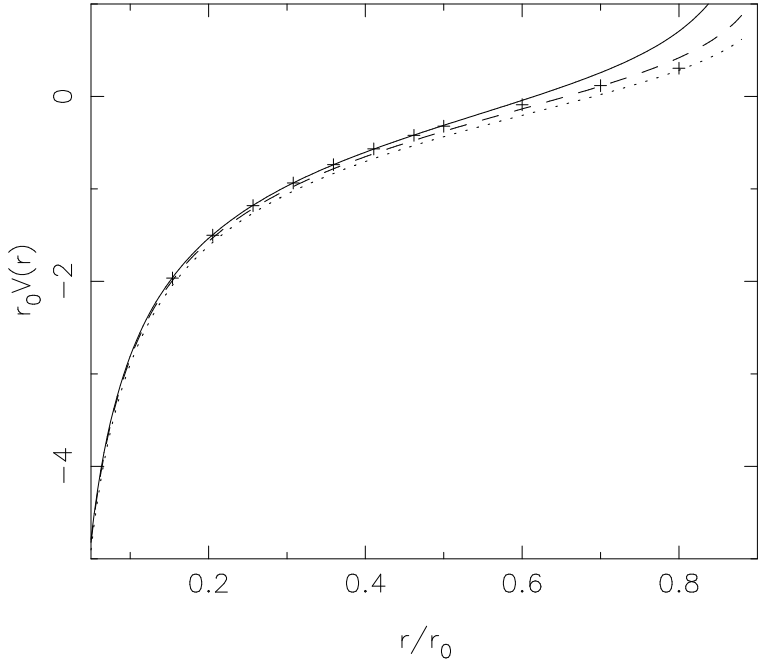}
\caption{\label{fig2} % \footnotesize
The lattice potential vs the Borel resummed potential
using $\tilde V_{0,2}$ (dotted), $\tilde V_{1,2}$(dashed),
and $\tilde V_{2,2}$ (solid).}
\end{figure}

The numerical integration of the Borel integral in Eq. (\ref{e5}) can be 
done easily in  $w$ plane defined by the mapping (\ref{e21}).
Using the Cauchy's theorem, the integration contour, for example,
on the upper half plane
in $w$ plane can be deformed to a ray off the origin 
to the unit circle in the
first quadrant. This trick allows to avoid the renormalon
singularity on the integration contour, and makes
the computation easy. For details we refer to \cite{lee4}.

For comparison with lattice
calculation we take the accurate data of the recent computation
employing large lattices \cite{lattice1}.
All the dimensional quantities are in units of 
the Sommer scale $r_0$ $(\approx 0.5 \,{\rm fm})$ \cite{sommer3},
where $r_0$ in terms of $\Lambda_{\overline{\rm MS}}$ ($\approx
238$\,MeV) is determined
in lattice computation \cite{alphagroup} to be
\be
r_0\Lambda_{\overline{ \rm MS}}=0.602(48) \,.
\ee
On the side of the perturbative potential, 
the Borel integration in Eq. (\ref{e5}) was done using the 
Borel transform  $\tilde V_{0,\rm 2},
\tilde V_{1,\rm 2},$ and $\tilde V_{2,\rm 2}$ in the bilocal expansion
(\ref{e8}).
The coupling constant $\alpha_s(1/r)$ was computed by numerically solving
Eq. (\ref{lambdams}) employing the four loop $\beta$  function
\cite{betafunction}.
Because of the divergent quark self energy
the lattice potential is determined only up to an $r$ independent
constant, so we subtracted such a constant from the lattice data
so that the lattice potential and the NNLO perturbative potential
agree exactly at $r/r_0=0.30798$.

The result is in Fig. \ref{fig2}.
Notice the rapid convergence of the resummed potential
at distances $r\alt 0.6r_0$ $[\approx (660$MeV)$^{-1}]$,
and the excellent
agreement of the NNLO potential with the lattice data.
The potential at leading order already 
fits the lattice values quite well.
It is remarkable that perturbative QCD is applicable at
distances as large as $r=(660 {\rm MeV})^{-1}$.

\section{Discussion and Summary}

The first thing we can learn form our result is that 
in the static potential
the leading renormalon is overwhelmingly dominant at short distances
and there cannot be any significant nonperturbative
effect other than that caused by the renormalon.
As already observed in \cite{sommer1,pineda1}, large linear
potentials at short distances  like those 
proposed in \cite{bali1,grz,simonov}
are excluded.

The rapid convergence of the perturbative potential in ABR
allows one to use the pole mass in perturbative
calculation of heavy quarkonium physics.
Because of the bad convergence of the truncated power series of the
static potential, there was a limit in 
the precision achievable with  perturbative QCD in
quarkonium physics \cite{sumino3,sumino4}. But,
it was soon realized that the cancellation of renormalons in
the pole mass and the static potential can be used to alleviate the 
problem \cite{hoang,beneke-98}.
Instead of using the pole mass directly,
one can achieve an improved convergence
by simultaneously expanding the
pole mass and static potential in the heavy quark Hamiltonian
in terms of the running coupling 
$\alpha_s(\mu)$ and  a short distance mass like 
the $\overline{\rm MS}$ mass \cite{sumino-pt1,sumino-pt2}.
Although this approach {\it avoids}
the renormalon problem, there could be
large logs in the perturbative expansion which could in principle
spoil the convergence.
Since the expansion involves two far-separated scales,
the heavy quark mass and $1/r$ ($\approx mv$,
where $v$ is the heavy quark velocity) large logs
like $\ln(r\mu)$ and/or $\ln(m/\mu)$ could survive for any choice of  $\mu$,
which in practice is typically taken values in-between the two scales.
With our resummation of the static potential, the convergence problem
at short distance 
is solved, so the pole mass needs not be abandoned in favor of a short
distance mass.
Once the pole mass is extracted by comparing, say, a calculated
quarkonium spectrum to an experimental value, the $\overline{\rm MS}$
mass can be obtained from the pole mass
by  resumming  the quark mass
expansion (\ref{e14}) in ABR.
Since the renormalon in the pole mass is essentially
same as that in the static potential, we 
can expect a rapid convergence of
the Borel resummation of the mass expansion, and we have checked
that this is indeed the case. As an example, for the bottom
quark ($N_f=4$) with
$\alpha_s(m_{\overline{\rm MS}})=0.22$ the `Borel resummed (BR)'
mass $m_{\rm BR}$,
which is defined as the real part of the integral term in Eq. (\ref{e15}), 
converges as 
\be
m_{\rm BR}=m_{\overline{\rm MS}}
(1+0.15769+0.00409-0.00028) \,.
\ee
Notice the rapid convergence.
The renormalon caused  nonperturbative effect 
$m_{\rm NP}$ in Eq. (\ref{e15})
can be determined in ABR, and its real part 
equals to $-{\rm Re}[V_{\rm NP}]/2$ that is given
in Eq. (\ref{nonp_v}).
An obvious advantage of the direct resummation of the renormalons 
is the separation of scales; The perturbative expansions for
the pole mass and the static potential are
resummed at their optimal scales $\mu=m_{\overline {\rm MS}}$ and
$\mu=1/r$, respectively, and there is no mixing of these scales 
as in the above implementation of renormalon cancellation using a
short distance mass. The absence of large logs and the excellent 
convergence of the resummed mass and potential are expected to provide
a new level of precision calculation for heavy quarkonium.

\begin{figure}
 \includegraphics[angle=-90 , width=10cm
 ]{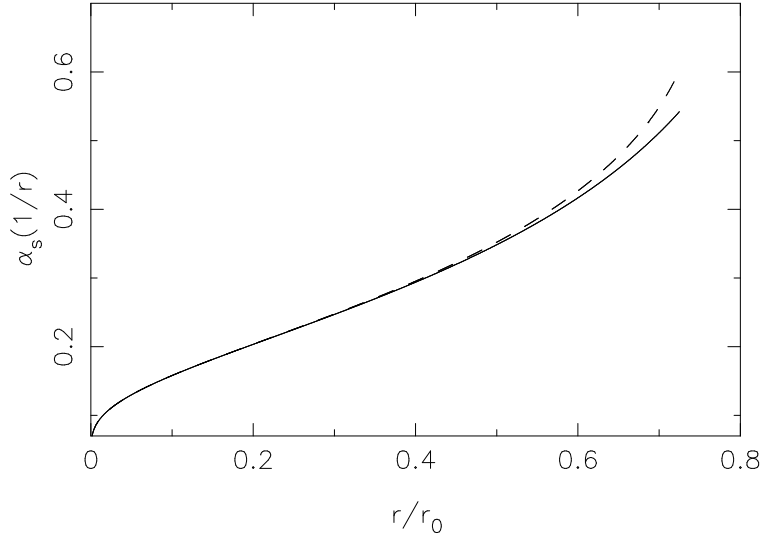}
\caption{\label{fig3} % \footnotesize
The strong couplings obtained by employing the four loop $\beta$ function
(solid)  and
its [2/3] Pad\'e approximant (dashed).}
\end{figure}

It is worthwhile to mention that the nonperturbative
effects $V_{\rm NP}$ and $m_{\rm NP}$ may actually decouple completely from
the quarkonium system. The renormalon cancellation between the pole mass
and the static potential means that the ambiguous imaginary parts in
these quantities cancel without the introduction of the nonperturbative
effects. This implies that the nonperturbative effects are actually 
spurious, appearing only at an intermediate step in Borel resummation,
and physical observables are completely independent of them. Specifically,
we may write the Hamiltonian of a heavy quarkonium
system as 
\be
H=2m_{\rm pole} +\frac{\roarrow{p}^2}{m_{\rm pole}} +V[r,\alpha_s(1/r)]\,.
\ee
Putting
\bear
m_{\rm pole}&=& m_{\rm BR}[m_{\overline {\rm MS}},\alpha_s(m_{\overline {\rm
MS}})]+{\rm Re} [ m_{\rm NP}]\,, \nonumber \\
V[r,\alpha_s(1/r)]&=&V_{\rm BR}[r,\alpha_s(1/r)] +{\rm Re} [ V_{\rm
NP}]\,,
\eear
where the BR potential $V_{\rm BR}$ denotes the real part of the
integral term in Eq. (\ref{e1}),
and using the cancellation of $2{\rm Re} [m_{\rm NP}]$ with 
${\rm Re}[V_{\rm NP}]$ in 
ABR,\footnote{This cancellation is not automatic but a
feature of ABR.}
we can write $H$ in terms of the BR quantities only: 
\bear
H= 2m_{\rm BR} + \frac{\roarrow{p}^2}{m_{\rm BR}}
+V_{\rm BR}[r,\alpha_s(1/r)] +
O(\roarrow{p}^2{\rm Re} [m_{\rm NP}]/m_{\rm BR}^2)\,.
\label{hamiltonian}
\eear
The remaining dependence on the nonperturbative effect suppressed by
an inverse power of the quark mass  is expected to cancel when 
higher order terms in quark mass expansion of the Hamiltonian
are taken into account.
This shows that the Hamiltonian in BR scheme is formally same
as that in the
on-shell scheme with the on-shell quantities 
$m_{\rm pole}$ and $V(r)$
replaced by the corresponding BR quantities.
Thus for physical
observables the specific form of the nonperturbative effects 
are not necessary.

The perturbative potential and the lattice  values in Fig. \ref{fig2}
begin to deviate
at $r\approx 0.6 r_0$, which we
regard as the failure of the perturbative potential at these distances.
It is interesting to observe that this deviation occurs approximately
at the same position where the four loop $\beta$ function fails.
The couplings $\alpha_s(1/r)$ obtained by running with the four loop $\beta$ 
function and its $[2/3]$ Pad\'e approximant, which differs from the former
only at orders higher than four loop,
are plotted in Fig. \ref{fig3}.
Notice that they begin to deviate approximately at the same distance
where the perturbative potential begins to fail. At $r= 0.6r_0$
[$\alpha_s(1/0.6r_0)=0.417$]
the $\beta$ function has the expansion
\be
\beta =-0.152 (1+0.308+0.143+0.097+{\ldots} )
\ee
which shows the convergence is quite slow at this distance.
It seems the coupling  grows too fast at these distances,
since a more slowly growing coupling would fit the lattice data.
This simultaneous deviations could be a coincidence,
but a more plausible
explanation would be that the failure of the $\beta$ function at these
distances results in an  unreliable coupling,
which then causes the deviation.
The $\beta$ function would not be all that
fails the perturbative potential.
Since there is a renormalon singularity at $b=3/2$ the bilocal
expansion (\ref{e8}) at a finite order
would certainly fail around $b\agt3/2$.
This does not cause any serious problem at small couplings, but
as the coupling increases this becomes problematic because
the Borel integral in Eq. (\ref{e5}) receives a sizable contribution
from the region far from the origin.  
By varying the upper bound of the
integration in Eq. (\ref{e5}) one can  easily check 
that the resummed potential at $r\agt 0.6r_0$
is indeed sensitive on the Borel transform at $b\agt 3/2$. 
This argument suggests that the applicability of the
Borel resummed perturbative potential could be extended 
to larger distances once 
we have a better control over the $\beta$ function and the Borel
transform at such distances.

Lastly, we note that the convergence problem of the 
truncated power series in the
perturbative potential is only one example, although a very
conspicuous one, of the problem of the QCD 
expansions in general, especially, at
low energies of a few GeVs. The problem was not so visible
in these expansions, since many were considered at a fixed scale,
not like the perturbative potential considered here where a
continuum of scale is involved.
Conventionally, in the OPE approach,
in these low energy expansions the physical quantity
is organized as the sum of a truncated power series and 
power corrections. Any difference between the truncated power series
and the (unknown) true value is swept over to 
the power corrections. Clearly, this approach fails in the
static potential  because the potential of the OPE approach
is just  the truncated power series plus an $r$ independent constant, which
we know has a bad convergence and disagrees with the lattice calculation. 
As already discussed more extensively in the Gross-Llewellyn Smith 
sum rule \cite{lee6} the solution to the problem is
the Borel resummation that properly accounts for the renormalon.
Without Borel resummation the bad convergence in the 
truncated power series results in wide fluctuations in the
power corrections as the order of perturbation varies, which is 
observed in many cases. See  \cite{yb,giz} for some examples. 

To summarize, we have shown that the Borel resummation with a proper
account of the renormalon singularity in the Borel plane 
can resolve the convergence problem of the perturbative static
potential and the pole mass, and the potential obtained in such a way is in
an excellent agreement with the lattice calculation.
Consequently, any significant nonperturbative effect at short distance 
other than the renormalon effect is excluded, and  
the pole mass can be used in perturbative calculation
of heavy quarkonium physics. The advantages of the direct resummation
of the renormalons include rapid convergence of 
the summations and 
absence of large logs, and these can open a new level of precision 
calculation for heavy quarkonium.
We also calculated in the framework of ABR the renormalon caused
nonperturbative effects in the static potential and the pole mass.
The resummation method developed here may be applied to the computation
of  heavy quarkonium spectra in an approach similar to 
that employed in \cite{sumino-spect}, where
the perturbative potential at short distance is combined with
the phenomenological potential at large distance.
Also it may be employed in the top threshold production.

\begin{acknowledgements}
The author is thankful to A.~Pineda for many helpful communications
and also to G.~Bali for a correspondence. This work was 
supported in part by BK21 Core Project.

\end{acknowledgements}

%\bibliographystyle{apsrev}
%\bibliography{pot}

\end{document}